\documentclass[a4paper]{jpconf}
\usepackage{amsmath}
\usepackage{graphicx}
\usepackage{citesort}

\begin{document}
\title{Study on Cooling of Positronium for Bose-Einstein Condensation}

\author{K. Shu$^{1,*}$, X. Fan$^{1,*}$, T. Yamazaki$^2$, T. Namba$^2$, S. Asai$^1$,\\ K. Yoshioka$^1$, M. Kuwata-Gonokami$^1$}

\address{$^1$Department of Physics, Graduate School of Science, The University of Tokyo, 7-3-1 Hongo, Bunkyo-ku, Tokyo 113-0033, Japan}
\address{$^2$International Center for Elementary Particle Physics (ICEPP), The University of Tokyo, 7-3-1 Hongo, Bunkyo-ku, Tokyo 113-0033, Japan}
\address{$^*$Corresponding authors}

\ead{\mailto{kshu@icepp.s.u-tokyo.ac.jp}(K. Shu), \mailto{xfan@icepp.s.u-tokyo.ac.jp}(X. Fan)}

\begin{abstract}
A new method of cooling positronium down is proposed to realize Bose-Einstein condensation of positronium. We perform detail studies about three processes (1) thermalization processes between positronium and silica walls of a cavity, (2) Ps\,-\,Ps scatterings and (3) Laser cooling. The thermalization process is shown to be not sufficient for BEC. Ps\,-\,Ps collision is also shown to make a big effect on the cooling performance. We combine both methods and establish an efficient cooling for BEC. We also propose a new optical laser system for the cooling.
\end{abstract}


\section{Introduction}
Positronium (Ps), a bound state of an electron and a positron, is the lightest atom, whose mass $m_{\mathrm{Ps}}=$1022\,keV. The two ground states of Ps, the triplet state~($1^{3}S_{1}$) and the singlet state~($1^{1}S_{0}$), are known as \textit{ortho}-positronium~(\textit{o}-Ps) and \textit{para}-positronium~(\textit{p}-Ps) respectively. \textit{Ortho}-positronium decays slowly into three photons with a lifetime of 142\,ns\cite{life}. On the other hand, \textit{p}-Ps quickly decays into two photons. The two states can be mixed with a weak magnetic field. Ps is therefore a good source of 511\,keV $\gamma$ ray. Furthermore Ps is a good tool to probe the gravity of anti-particle, since Ps is a purely particle and anti-particle system.

Bose-Einstein condensation (BEC) is one of the most interesting phenomena of the quantum physics. Behavior of quantum particles can be magnified into a macroscopic level to be directly observed in BEC state, and BEC provides various applications. The first observation of BEC of weakly interacting bosonic atomic gas was found using $^{87} \mathrm{Rb}$ gas in 1995\cite{RbBEC} and opened a new era of studying the macroscopic behavior of quantum gas. Ps BEC is very attractive since it would provide a 511\,keV $\gamma$ ray laser\cite{Vanyashin1994,gammarayLaser1,gammarayLaser2} and macroscopic behavior of anti-particle gravity could be observed.

\begin{figure}
  \centering
  \includegraphics[width=7.5cm,clip]{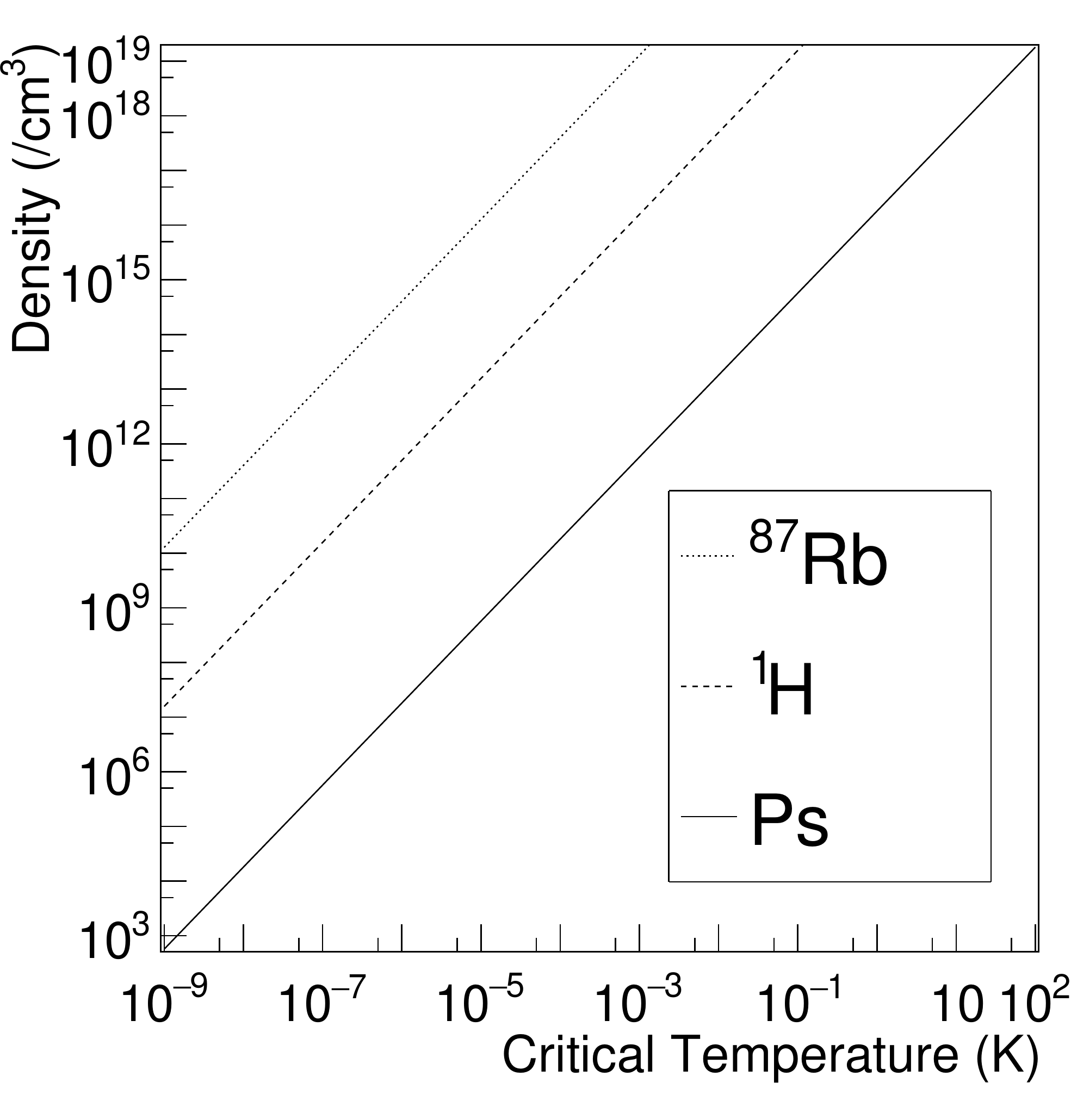}
  \caption{Relation between density and critical temperature of BEC transition. Left side of each line is a region where the atoms are in BEC phase.}
  \label{fg:BECTc}
\end{figure}

The de Broglie wave length, $\lambda_D$, and the density, $n$, play an important role for BEC. The critical temperature, $T_\mathrm{C}$, is determined by the following formula\cite{pethick2002bose}: 
\begin{equation}
  n \lambda_D^3 = n \left(\frac{2 \pi \hbar ^2}{m k_B T_C}\right)^{3/2} = 2.612,
\end{equation}
\centerline{$\hbar$: The reduced Planck constant, $m$: A mass of an atom, $k_B$: The Boltzmann constant.}
\ \\
Figure \ref{fg:BECTc} shows a relation between $T_\mathrm{C}$ and $n$, for $^{87}\mathrm{Rb}$, $^1$H and Ps. Since Ps is very light, BEC can be achieved at a few hundreds millikelvins for $n \sim 10^{15}\,\mathrm{/cm^3}$. This $T_\mathrm{C}$ is much higher than that for $^{87}\mathrm{Rb}$ and $^1$H. 

It is essential to rapidly cool dense Ps atoms down in order to achieve a high phase-space density. The conventional way to cool Ps down is creating Ps atoms in a cold material with voids into which Ps can escape. The most recent experiment could cool them down to 150\,K by using cold silicon nanochannels (less than 10\,nm in diameters)\cite{Mariazzi2010} in which Ps atoms were created and thermalized via many collisions with the channel's walls. This minimum temperature was limited by the sizes of voids. If a material with smaller voids is used, momentum exchanges between Ps and the walls will be more efficient. However, a quantum confinement effect which is remarkable due to the large de Broglie wave length of Ps prevents Ps atoms from being cold inside a too small void\cite{Mariazzi2008}. Cooling with a laser has also been proposed to achieve less than 10\,K of Ps atoms\cite{Liang1988,Kumita2002,Crivelli2014}. As for density, $2\times 10^{15}\,\mathrm{/cm^3}$ was achieved recently\cite{Cassidy2007} by a positron accumulator with $\mathrm{^{22}Na}$ radioisotope. Much effort is still necessary for both cooling and accumulation to achieve the phase-space density for BEC phase transition.

In order to realize BEC of Ps atoms, using a cold silicon cavity whose dimension was around 1\,$\mathrm{\mu m}$ was proposed for trapping and cooling down Ps by collisions between the walls and Ps\cite{Platzman1994}. However, we will later show that the cooling with the large cold cavity is insufficient. In this article, we propose a new cooling method using both the cold cavity and the laser cooling. The efficiency of the cooling is estimated taking into account effects of dense Ps atoms.

\section{Setup}
\label{sc:Setup}
\begin{figure}
  \centering
  \includegraphics[width=13cm]{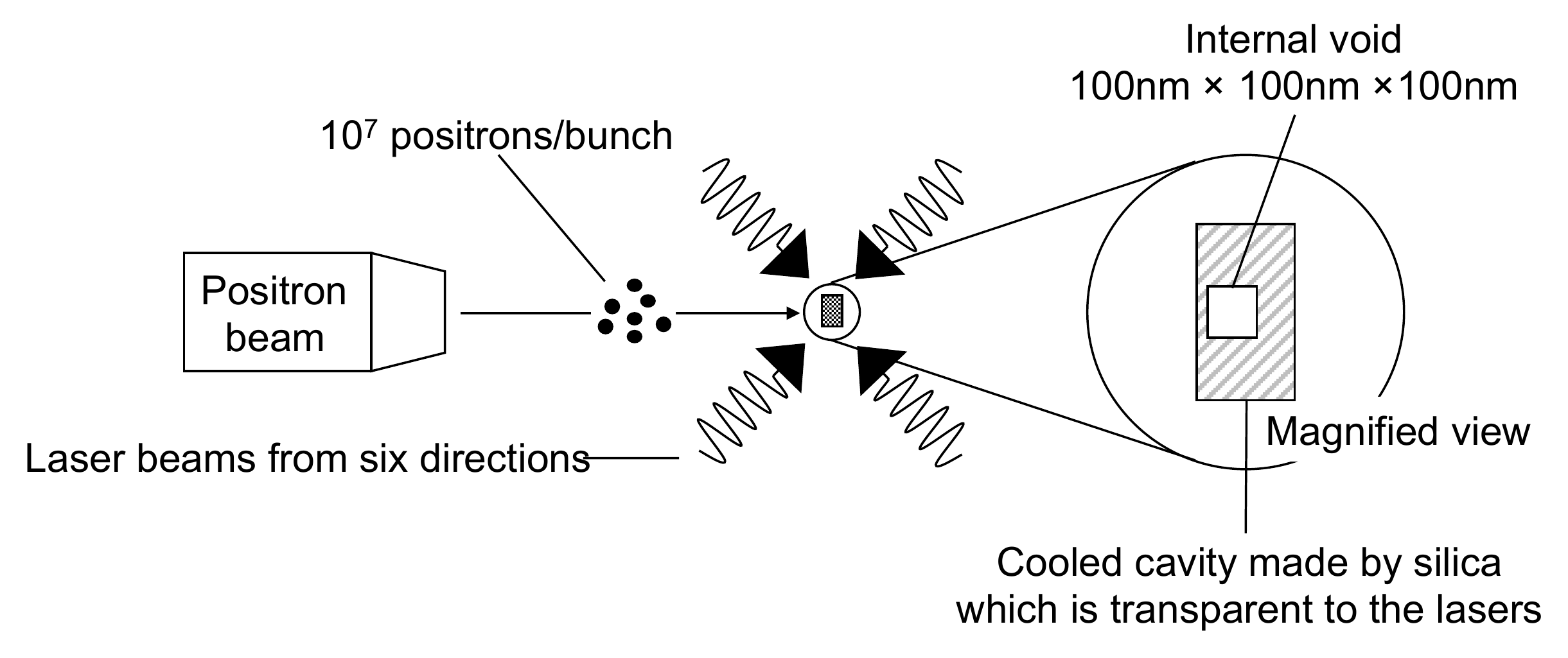}
  \caption{The schematic diagram of the experimental setup. The two laser beams which are perpendicular to the sheet are not written.}
  \label{fg:Setup}
\end{figure}

A conceptional view of our experimental setup is shown in Fig. \ref{fg:Setup}.
Positrons are stored in a small trapping cavity made by silica. A 5\,keV positron beam of $10^7$ positrons per bunch will be used. This number of positrons is already possible elsewhere\cite{Cassidy2006}. The positrons are focused and injected into the cavity which is cooled at 1\,K. The cavity has an internal void whose dimension is assumed to be a cube of $100\,\mathrm{nm} \times 100\,\mathrm{nm} \times 100\,\mathrm{nm}$. About $4\times 10^3$ fully spin-polarized Ps atoms are assumed to be left inside the void, which in turn means $n=4\times 10^{18}\,\mathrm{/cm^3}$, while the focusing technique and overall efficiency of the conversion from the positrons to Ps have to be studied in future. Ps atoms have an initial kinetic energy of around 0.8\,eV\cite{Nagashima1995} and are confined inside the cavity. The cavity is irradiated by UV laser beams which are configured as optical molasses: by three orthogonal pairs of counter-propagating beams. The laser photons can go through into the cavity because silica is transparent to this light. The laser system is described in section \ref{sc:LaserImplementaion}. 

In this setup, Ps atoms have the following interactions:
\begin{itemize}
  \item Thermalization by interactions with silica walls of the cold cavity,
  \item Ps\,-\,Ps two-body interactions,
  \item Cooling by optical transitions and momentum recoils by photons.
\end{itemize}
As discussed in the following sections, the cooling through the thermalization process is efficient for Ps atoms with high energy while the opposite for laser cooling. These two cooling processes are complementary, so we propose a new cooling scheme in two stages: initially by Ps\,-\,silica interactions and then by laser fields after the former becomes inefficient. We evaluate cooling efficiency of each process and see whether it is enough to achieve BEC transition. 

\section{Thermalization in the silica cavity}
\label{sc:Thermalization}
The thermalization process is evaluated by using the classical interaction model\cite{Nagashima1995}. This model presumes that thermalization will proceed through classical elastic collisions between Ps and grains of a surrounding material. According to the model, an average kinetic energy of Ps, $E$, evolves as follows:
\begin{equation}
  \frac{\mathrm{d}E}{\mathrm{d}t} = -\frac{2}{LM}\sqrt{2m_{\mathrm{Ps}}E}\left(E-\frac{3}{2}k_BT\right).
  \label{eq:Thermalization}
\end{equation}
Here $M$ is an effective mass of surrounding grains, $L$ is the mean free path of the collisions with grains and $T$ is temperature of a surrounding material. This model can well describe the thermalization process which were measured by various techniques\cite{Chang1987,Takada2000,Shibuya2013}.

\begin{figure}
    \centering
    \rotatebox{0}{\includegraphics[width=10cm,clip]{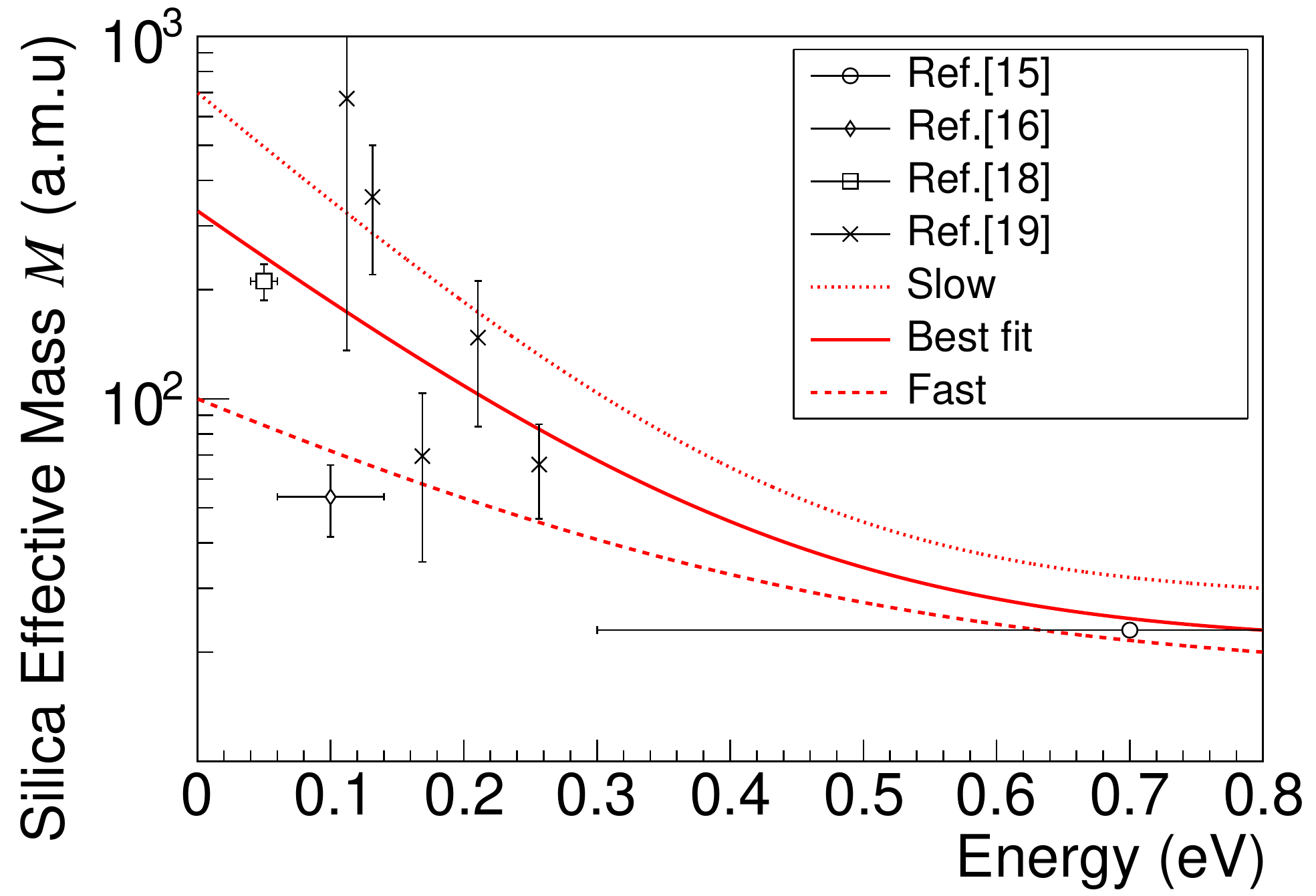}}
    \caption{Effective mass of a silica grain versus kinetic energy of Ps. Error bars for the vertical axis are measurement errors and ones for the horizontal axis are measured energy regions if there are. Lines are the fitting results with our estimated function (see text).}
    \label{fg:EffectiveM}
\end{figure}

Experimental results are used in order to determine $M$. Various results of the measurements\cite{Nagashima1995,Shibuya2013,Chang1987,Nagashima1998} are shown in Fig. \ref{fg:EffectiveM}.
$M$ is shown as a function of Ps kinetic energy in which it was measured, because the effective mass could depend on kinetic energy of interacting Ps as suggested\cite{Nagashima1998}. This is because the number of phonon modes which Ps can excite at collisions decreases as kinetic energy of Ps does. We estimate energy dependence of $M$ in a.m.u as $M=21+308\,\exp{\left(-\frac{E}{0.16\,\mathrm{eV}}\right)}$ in order to reproduce the experimental results as shown in Fig. \ref{fg:EffectiveM} by the solid line which is named as ``Best fit''. The uncertainty is large because precision of the measurements is still limited. The range of the predictions is also shown in Fig.\ref{fg:EffectiveM} as ``Fast'' and ``Slow''. 

In addition to the interaction with the silica walls, the Ps\,-\,Ps interactions must be taken into account in the case of high density. The main process is the elastic $s$-wave scattering of spin-polarized Ps atoms. This process leads Ps atoms to quasi thermal equilibrium: the energy distribution of atoms becomes the Maxwell-Boltzmann distribution with an approximation of classical scattering. The total cross section ($\sigma$) of the scattering is given by $\sigma=4\pi a^2$ where $a$ is a scattering length, $a=0.16\,\mathrm{nm}$\cite{Oda2001,Ivanov2002}. The mean free time ($\tau$) of scatterings depends on a number density of Ps ($n$), an average velocity ($\bar{v}$), and $\sigma$ as $\tau = 1/n \sigma \bar{v}$. This interval with $n=10^{18}\,\mathrm{/cm^3}$ is less than 100\,ps even at 10\,K. The quasi thermalization among Ps atoms is quite fast compared to the thermalization process between Ps atoms and the silica walls.

\begin{figure}
  \centering
  \rotatebox{0}{\includegraphics[width=10cm,clip]{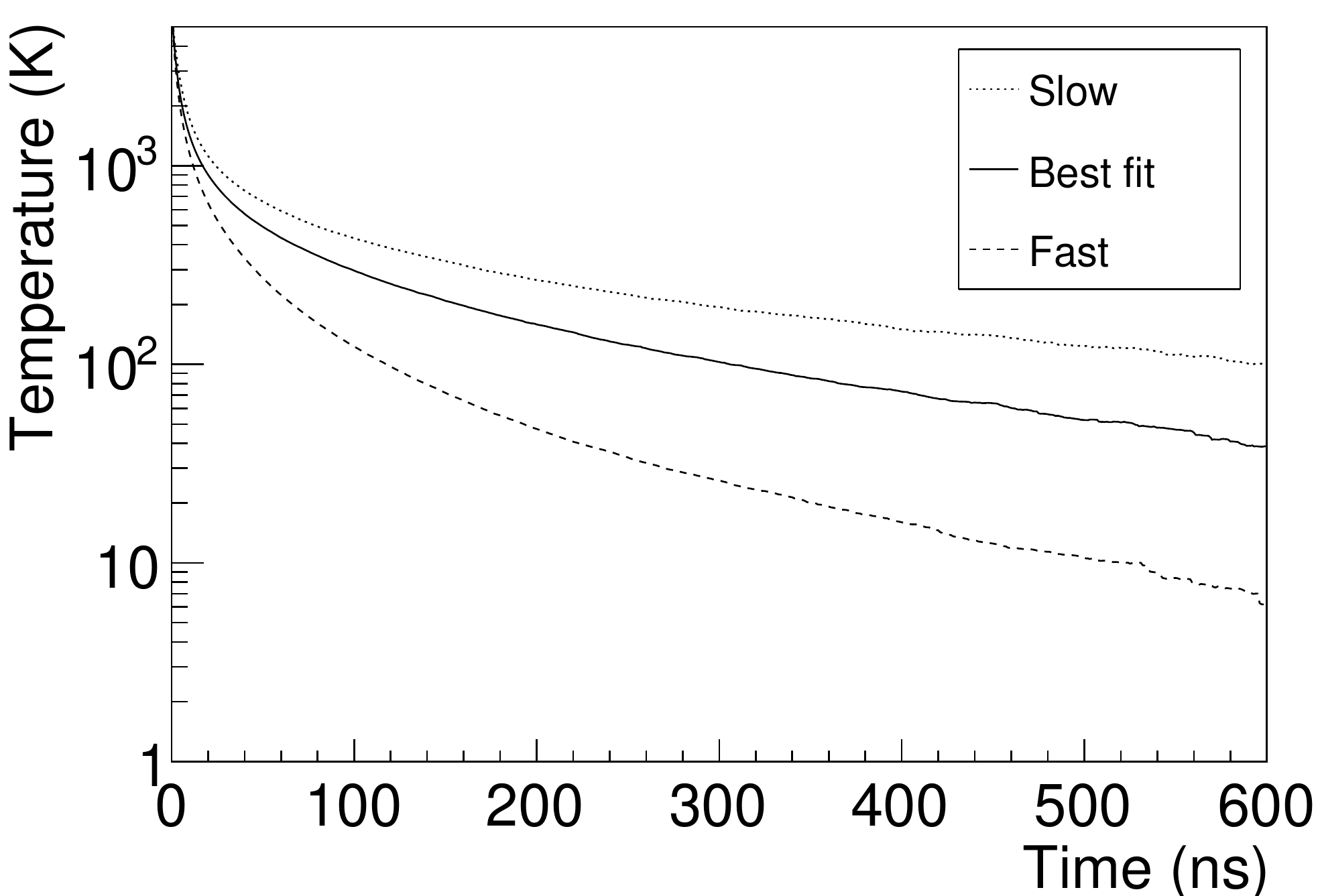}}
  \caption{Time evolutions of temperature in the silica cavity. The evaluations are performed with different three estimations of $M$. Legends show which estimation curve is used. ``Slow'', ``Best fit'' and ``Fast'' correspond to the curves in Fig. \ref{fg:EffectiveM}.} 
  \label{fg:Thermalization}
\end{figure}

The interactions with the silica wall and Ps\,-\,Ps elastic scatterings are simulated by a Monte Carlo method, which is explained in an appendix in detail. Time evolutions of temperature are shown in Fig. \ref{fg:Thermalization}.
As for ``Best fit'' estimation of $M$, Ps can be cooled to 300\,K in 100\,ns from the initial energy of 0.8\,eV. The further cooling is so slow that it takes 500\,ns to reach 100\,K even though the cavity is at 1\,K. Our estimation is consistent with previous studies in which Ps would not thermalize at low temperatures\cite{Kiefl1983,Saito1999,SurkoSaito2001}. 

However the thermalization process strongly depends on the effective mass of a silica grain as shown in Fig. \ref{fg:Thermalization}.  It is necessary to perform additional measurements precisely in order to determine $M$. 

There is another interaction model with silica\cite{Mariazzi2008,Morandi2014}. In this model, Ps atoms emit or absorb acoustic phonons on the cavity walls through deformation potential scatterings. The interaction rates can be deduced by the first-order perturbation theory. We also evaluate the thermalization process in this model by fitting to reproduce the experimental results\cite{Chang1987,Takada2000,Shibuya2013}. There is no observable difference between the two models for Ps atoms of higher than around 300\,K. At lower energy, the thermalization process in this phonon model is more insufficient than in the classical model because of less stimulated emission at low temperatures in the phonon model. In our cooling scheme, the thermalization process is crucial only at the initial stage of cooling to around 300\,K. The difference to the final result is therefore negligible.

\section{Laser Cooling}
\label{sc:LaserCooling}
\subsection{The evaluation of laser cooling}
It is necessary to accelerate the cooling down from a few hundreds of Kelvins. It is efficient to use $1s\leftrightarrow 2p$ transition for laser cooling of Ps because of a large recoil momentum by photons. The difference between these two energy levels corresponds to 243\,nm UV light.  

Optical transitions induced by the laser can be modeled with the rate equation approach\cite{Iijima2001} because the time scale of cooling down, more than 100\,ns, is much longer than the time constant 3.2\,ns of the spontaneous emission from $2p$ to $1s$. The stimulated transition rate, which is called as ``Einstein $B$ coefficient'', can be calculated by the flux of photons and the cross section of the interaction between Ps and resonant photons. This coefficient can be calculated as follows:
\begin{equation}
  B(t, \vec{x}, \vec{v}) = \int \mathrm{d}\omega\, \frac{I(t,\vec{x},\omega)}{\hbar \omega} \cdot \frac{4}{3}\pi^2 \alpha \omega_0 |X_{12}|^2 \cdot \frac{1}{2\pi} \frac{\Gamma/2}{(\omega(1-\vec{\hat{k}}\cdot \vec{v}/c)-\omega_0)^2+(\Gamma/2)^2},
  \label{eq:B}
\end{equation}
\centerline{$t$: time, $\vec{x}$: Position of Ps, $\vec{v}$: Velocity of Ps, $\vec{\hat{k}}$: Direction of laser photons,}
\centerline{$I(t,\vec{x},\omega)$: Intensity per frequency of the laser,}
\centerline{$X_{12}$: The matrix element of $1s$-$2p$ transition,}
\centerline{$\omega$: The frequency of interacting photons in the laser}
\centerline{$\omega_0$: The resonant frequency of $1s$-$2p$ transition.}
\ \\
Gaussian profiles are assumed for frequency/timing/space domains of the intensity. $\Gamma=313~\mathrm{MHz}$ is the natural line width so the last term in the equation (\ref{eq:B}) represents the Breit-Wigner line shape including the first-order Doppler effect. The internal state evolves according to this stimulated transition rate and the spontaneous emission rate from $2p$ state. A Ps is recoiled by $\hbar\omega/m_{\mathrm{Ps}}c\simeq 1.5\times 10^3~\mathrm{m/s}$ when it emits or absorbs a photon. The direction of the recoil is random for spontaneous emission while for stimulated absorption/emission it is the same as photons in the laser. An important feature is that the annihilation to gamma-rays from $2p$ state is very slow (10\,$\mathrm{s^{-1}}$) compared to that from $1s$ state. This means that the maximum excitation effectively increases the lifetime of Ps by twice to 284\,ns. 

\begin{figure}
    \centering
    \includegraphics[width=10cm,clip]{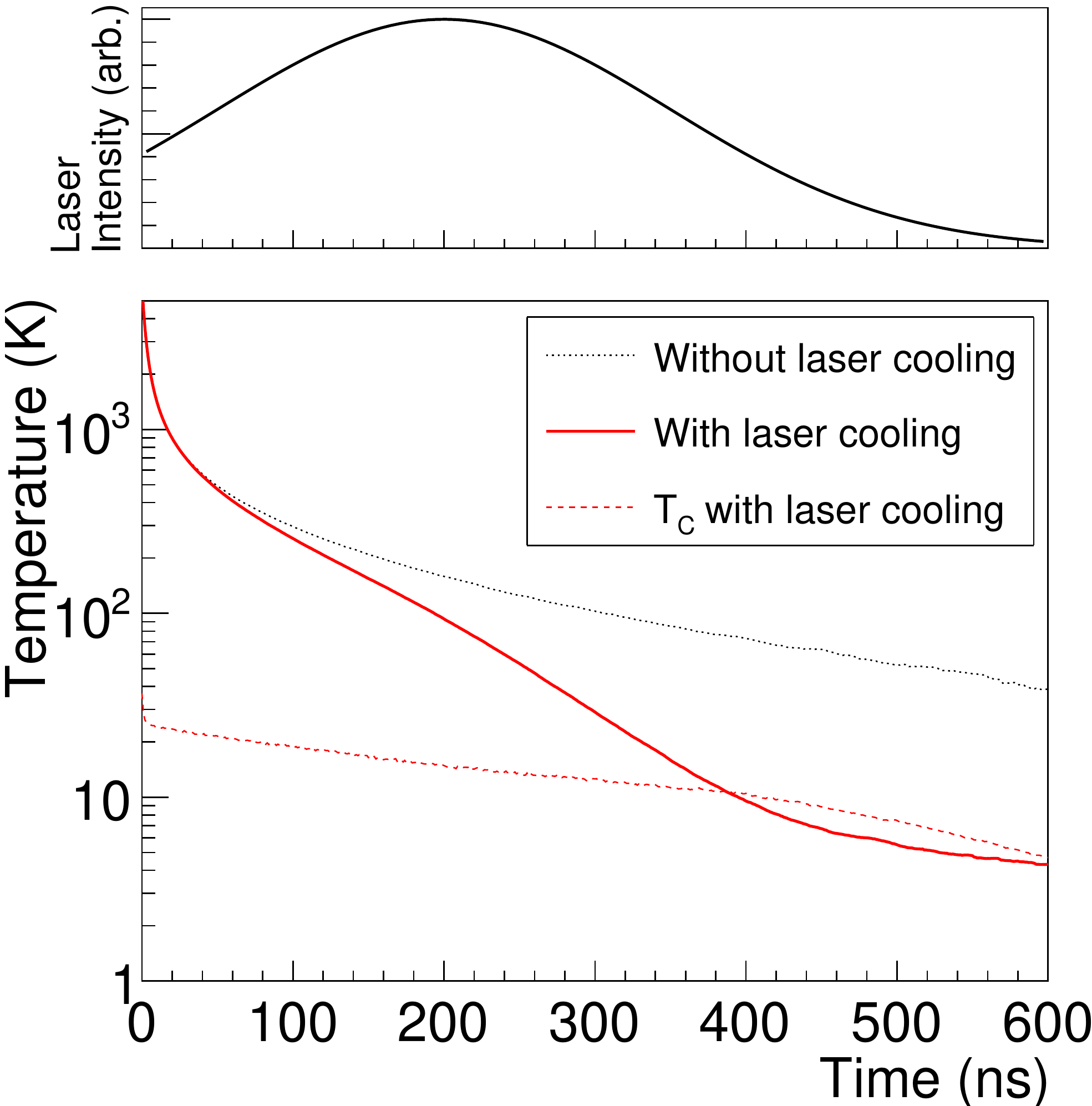}
    \caption{The upper part: Irradiated laser intensity in an arbitrary unit versus time. The lower part: Time evolutions of temperature and $T_\mathrm{C}$. $T_\mathrm{C}$ decreases as the density of $1s$ Ps does due to the annihilation.}
    \label{fg:LaserCooledTemperature}
\end{figure}

Table \ref{tb:LaserParameters} shows a summary of laser parameters.
\begin{table}
  \centering
  \caption{The summary of laser parameters of the 243\,nm laser system.}
  \begin{tabular}{cr}
    \hline
    Parameter Name & Value \\
    \hline\hline
    Pulse energy & 40\,$\mathrm{\mu J}$ \\
    Center frequency & 1.23\,PHz$-\Delta(t)$ \\
    Frequency detune $\Delta(t)$ & $\Delta$(0\,ns)=300\,GHz \\
     & $\Delta$(300\,ns)=240\,GHz \\
    Bandwidth\,(2$\sigma$) & 140\,GHz \\
    Time duration\,(2$\sigma$) & 300\,ns \\
    Beam size\,(2$\sigma$) & 200\,$\mathrm{\mu m}$ \\
    \hline
  \end{tabular}
  \label{tb:LaserParameters}
\end{table}
The laser is a pulsed laser and its energy is 40\,$\mu$J, which is divided into the six beams and focused into 200\,$\mu$m at the cavity. The timing of the peak intensity is delayed by 200\,ns from the creation of Ps atoms as shown in the upper part of Fig. \ref{fg:LaserCooledTemperature}. The time duration is 300\,ns in order to cover the long duration necessary for cooling. The center frequency of the laser field is detuned from 1.23\,PHz which corresponds to 243\,nm wavelength. This detune, $\Delta(t)$, is 300\,GHz at the beginning and then up-chirped to 240\,GHz in order to compensate the decrease of Ps velocities. The bandwidth around the center frequency is 140\,GHz in order to excite Ps atoms with a wide range of velocities. The required frequency chirp and bandwidth are quite large compared to standard systems for cooling other atoms. It is because of the large Doppler shift of Ps due to its light mass compared to any other atoms. The laser system with these features is a challenge.

\begin{figure}
  \centering
  \rotatebox{0}{\includegraphics[width=10cm,clip]{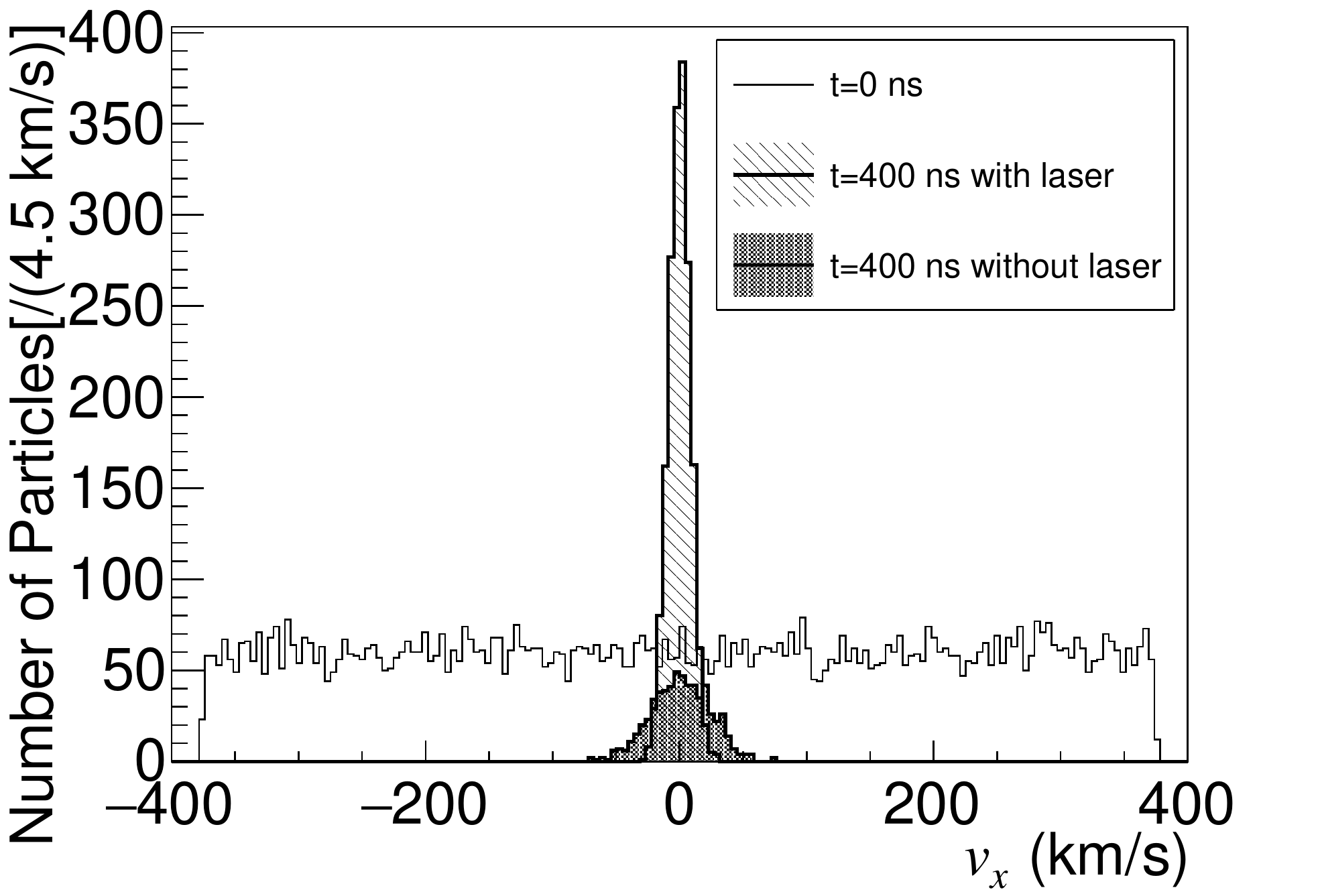}}
  \caption{Distributions of Ps velocity with and without laser cooling. The horizontal axis represents $v_x$ which is one component of Ps velocity vector. The number of remained atoms with laser is larger than that without laser because of the extended lifetime of the annihilation by excitations to $2p$ state.}

  \label{fg:VelocityDistribution}
\end{figure}

\begin{figure}
  \centering
  \rotatebox{0}{\includegraphics[width=10cm,clip]{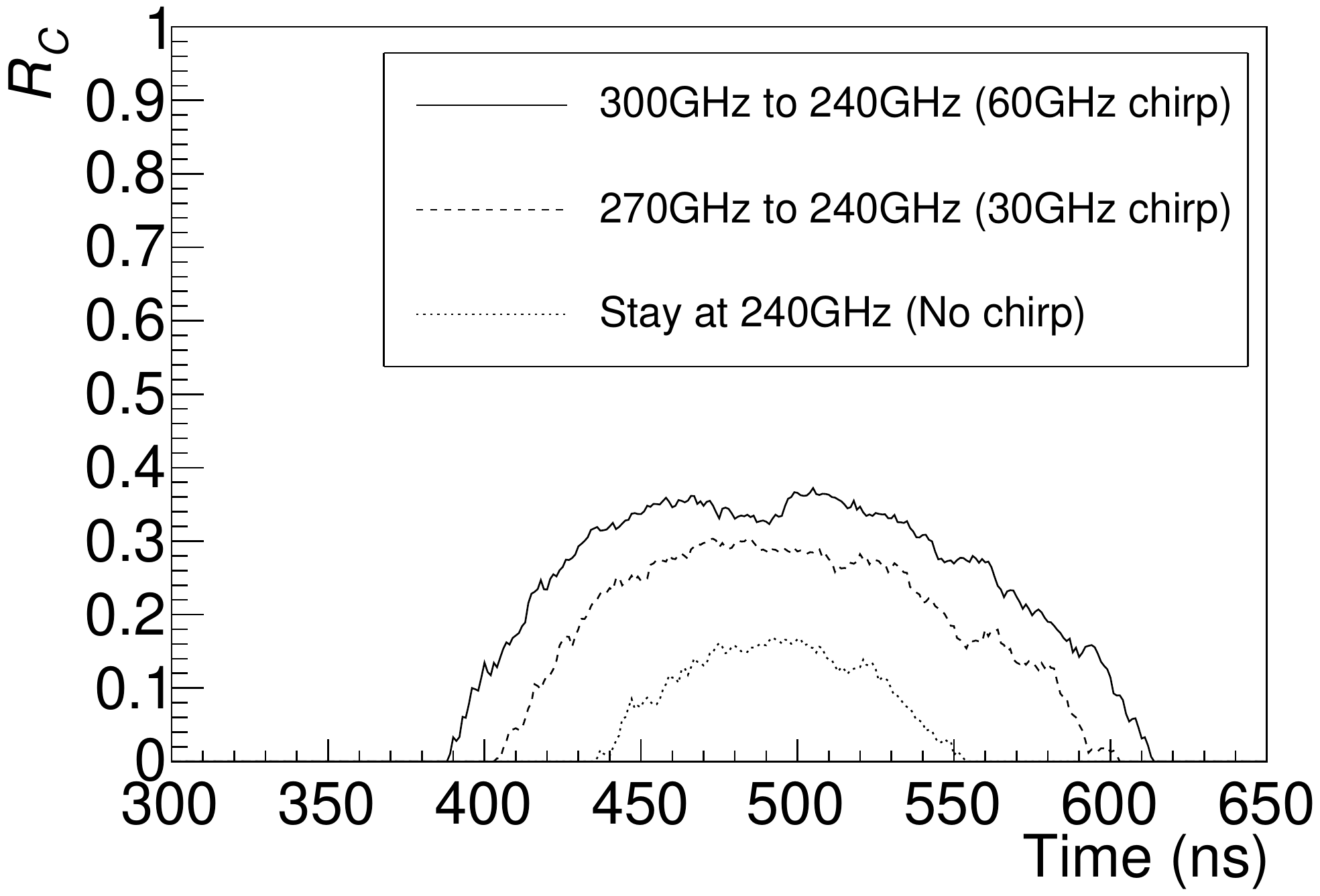}}
  \caption{Condensate fractions, $R_C$, with different frequency chirps versus time. $\Delta(300\,\mathrm{ns})$ is fixed at 240\,GHz and chirp ranges are 60\,GHz, 30\,GHz and 0\,GHz. Fractions increase in the beginning due to the cooling and then decrease due to the annihilation of Ps.}
  \label{fg:CondensateFraction}
\end{figure}

The cooling effect by the laser is evaluated by another Monte Carlo simulation, which includes laser effects with the thermalization and the Ps-Ps two-body scatterings. The details are also given in the appendix. ``Best fit'' estimation of $M$ in Fig. \ref{fg:EffectiveM} is used for the thermalization process. Distributions of Ps velocity at different times are shown in Fig. \ref{fg:VelocityDistribution}.
The distributions quickly become Maxwell-Boltzmann distributions by Ps\,-\,Ps scatterings. This means that Ps atoms are always in quasi thermal equilibrium and have well-defined temperature which can be calculated by a width of a velocity distribution. The number of remained atoms is increased with the laser because of the longer lifetime as described before. The time evolution of temperature and $T_\mathrm{C}$ are shown in the lower part of Fig. \ref{fg:LaserCooledTemperature}.
$T_\mathrm{C}$ is calculated from the density of Ps in $1s$ state. The cooling effect by the laser becomes dominant below several hundreds of Kelvins. After around 400\,ns, the temperature becomes less than $T_\mathrm{C}$. This means that the phase transition to BEC can be achieved by our cooling scheme. Figure \ref{fg:CondensateFraction} shows condensate fractions over remained atoms, $R_C=1-\left( T/T_C \right)^{\frac{3}{2}}$\cite{pethick2002bose}, which are calculated with an assumption that Ps atoms are non-interacting bosonic systems. 
More than 30\% of the remained atoms will be in the condensate with the laser system described above. 

A gain of the chirp is demonstrated in Fig. \ref{fg:CondensateFraction}, in which $R_C$ are calculated with $\Delta(0\,\mathrm{ns})$ being changed to 270\,GHz or 240\,GHz while $\Delta(300\,\mathrm{ns})$ are fixed at 240\,GHz. $R_C$ without chirp is only around 0.1 and the time interval of condensation is shortened to less than 100\,ns. Even 30\,GHz chirp can increase the fraction by twice than that without chirp. Therefore, 30\,-\,60\,GHz chirp is enough for the efficient cooling.

\subsection{Implementation of the laser system}
\label{sc:LaserImplementaion}
\begin{figure}
  \centering
  \includegraphics[width=15cm]{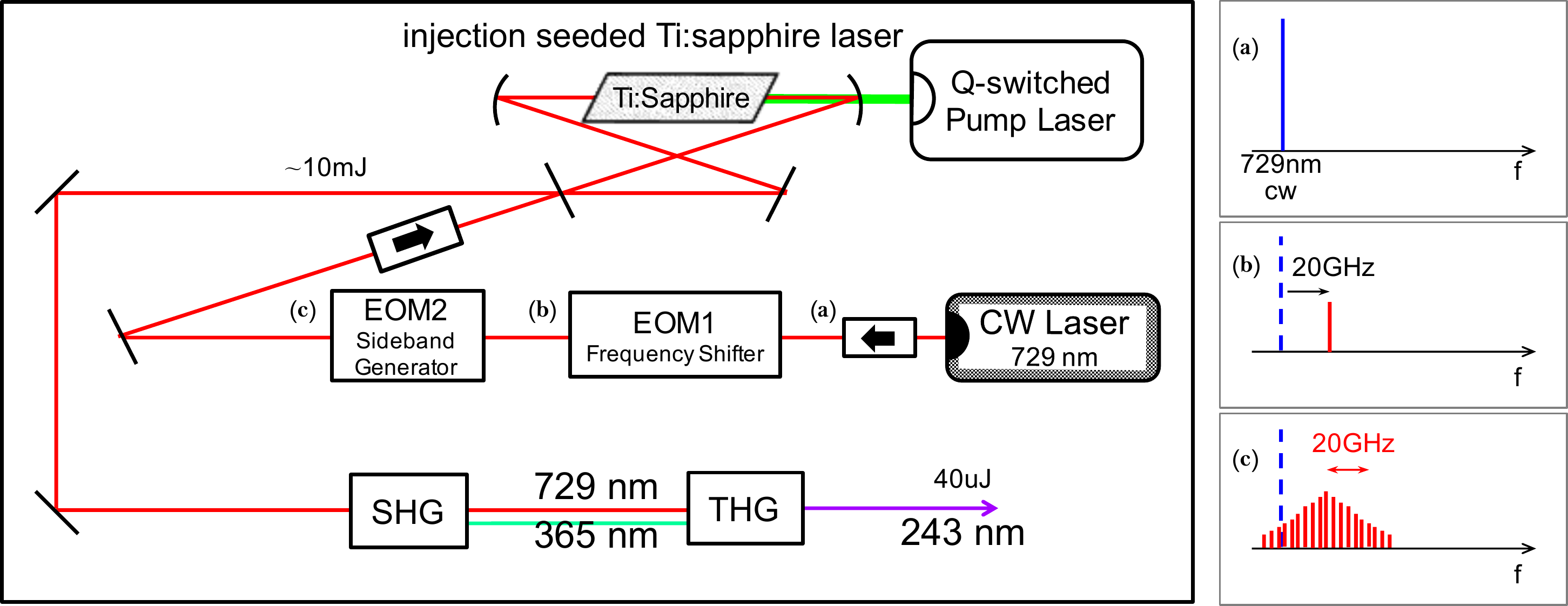}
  \caption{The conceptual diagram of our designed laser system. The 729 nm single mode CW light is frequency-controlled at EOMs. Pulse shaping and amplification are conducted at the following injection seeded Ti:Sapphire laser. Ti:sapphire crystal is pumped by Q-switched 532\,nm pulsed laser. The light is converted to 243 nm at the last SHG and THG part. The pictures on the right side show the frequency information of the 729\,nm light: (a) before entering EOMs, (b) ejected from EOM 1 which generates up-shift to 20\,GHz, (c) ejected from EOM 2 which generates sidebands of about 20\,GHz range. The frequency shift and broadening are multiplied by three at the last THG and the desired parameters in Table \ref{tb:LaserParameters} can be obtained.}
  \label{fg:laser}
\end{figure}

Though the specific parameters listed in Table \ref{tb:LaserParameters} are technically challenging, it can be achievable using various techniques which already exist. The technically challenging points are the large and fast frequency chirp with the optical amplification for the long time duration of more than 100\,ns. The basic idea is to use the third harmonics generation of 729\,nm light, whose frequency and pulse shape are precisely controlled. Figure \ref{fg:laser} shows a conceptual diagram of our designed laser system. 
A single mode 729\,nm CW laser is used as a master laser. The CW light is modulated at the electro-optic modulator 1, EOM 1, to generate a 20\,GHz sideband. This generated sideband is used as a center frequency of the frequency chirp. The advantage to use the sideband as a center frequency is that the chirp can be electrically controlled by changing modulation frequency applied to EOM 1. At EOM 2, 20\,GHz broad sidebands are generated around the first sideband. 1\,GHz modulation frequency to EOM with 20 sidebands generation is enough to cover all the Ps Doppler broadening. After the sideband generation, the seed light is injected into injection seeded Ti:sapphire laser. The gain of the laser can be controlled by adjusting the waveform of the pump laser and reflective indexes of an output coupler so the desired amplification along 300\,ns can be obtained. The pulse energy is about 10\,mJ. Now the amplified pulse is well controlled in both time domain and frequency domain, and enters the following Second Harmonics Generation, SHG and Third Harmonics Generation, THG, to achieve desired 243\,nm light. The 10\,mJ injected pulse energy is high enough to obtain 40\,$\mu$J energy. Furthermore, the 20\,GHz frequency shift and 20\,GHz broadening are also multiplied by three at THG, so the frequency parameters in Table \ref{tb:LaserParameters} can be obtained. The 243\,nm light will be sent to Ps generation cavity and used as cooling light. 

The repetition rate of the positron beam will be less than 10 Hz, so the assumed pulse energy is not so hard to achieve. Coincidence between the positron source and the laser system will also be easy by synchronising the positron system and the pulsed pump laser and EOMs. The most challenging part will be the large and fast frequency modulation at the two parts of EOM. In order to achieve this large frequency modulation, broadband traveling-wave type optical modulator will be used as a frequency shifter. Though this type of modulators have been mainly used in optical communication wavelength, modulation up to 100\,GHz also exists in 1064\,nm\cite{EOM}. We are developing this broadband device compatible with 729\,nm.

\section{The Roadmap for Ps BEC}
There are three steps to achieve Ps BEC. At first, the cooling process with silica should be confirmed. We are now measuring the thermalization process precisely. The second step is to develop the laser system. Some studies are ongoing with basic technology already developed. The last step is to develop the focus system of the slow positron beam. The beam should be focused into 100\,nm while it can currently be focused only into 25\,$\mu$m\cite{PositronMicroBeam} by different technique.

The detection of the transition to BEC will also be possible by the same technique as the precise measurement of the Ps thermalization. Another method could be using a characteristic spectrum of annihilation gamma-rays from the condensate. It is also under study.

\section*{Acknowledgements}
We would like to express sincere gratitude to Prof. Y. Nagashima, Prof. H. Saito, Dr. J. Omachi, and Mr. Y. Morita for helpful discussions. We warmly thank Dr. A. Ishida for his useful advice.

\section*{Appendix}
Details of the two Monte Carlo simulations are given in this appendix. For both simulations, $10^4$ atoms are created initially with monochromatic (0.8\,eV) and isotropic velocities. A velocity and an internal state of each atom are traced at the same time as a brute-force method. Time evolution of those quantities are divided into short time steps. Random numbers are generated at each step to compare with rates of interactions. Durations of the steps, which are typically 0.1\,ps, are determined so that all possible interactions happen with low probability ($<1\%$). Results shown in the text are obtained by one execution without any averaging. Some notes are as following:
\begin{itemize}
  \item Positions of the atoms are not traced because the trapping region of Ps, the 100\,nm cube, is narrow enough compared with the laser beam size. 
  \item The number of simulated atoms ($10^4$) is more than the assumed initial number, $4\times 10^3$, in order to decrease a statistical uncertainty of the simulation, but the rate of the Ps-Ps scatterings is scaled to represent the assumed condition in section \ref{sc:Setup}.
\end{itemize}
Following interactions are coded in the simulations.
\begin{description}
   \item[Thermalization through collisions with the silica cavity wall]\mbox{}\\
     A change of an average Ps kinetic energy is calculated by differential Eq. (2) with a linear approximation. For the parameters, $L=100\,\mathrm{nm}$ assuming $L=\sqrt[3]{V}$\cite{SurkoSaito2001} where $V$ is a volume of the trapping cavity and $M$ is determined by the curves in Fig. \ref{fg:EffectiveM}. A kinetic energy of each Ps atom is then evolved according to the result of the calculation for each time step whose duration is determined to suppress energy changes to be less than a percent for this interaction. 

 \item[Ps-Ps two-body interactions]\mbox{}\\
   Pairs of Ps atoms are made randomly and then momenta of final states are calculated as a result of the elastic $s$-wave scatterings for each pair. Rates of scatterings are calculated as $1/\tau(=n\sigma \bar{v})$, here $n$ is scaled to match the initial density ($4\times 10^{18}\,\mathrm{/cm^3}$).   

 \item[The annihilations and the spontaneous de-excitation of Ps]\mbox{}\\
   The annihilation of $1s$ Ps (142\,ns lifetime) is included, while the annihilation of $2p$ Ps is ignored because a time duration of the simulations, 600\,ns, is much shorter than its lifetime (100\,ms). After a Ps atom annihilates, the atom is not included in the simulation. The spontaneous de-excitation of $2p$ Ps into $1s$ Ps with 3.2\,ns time constant is also included.    

  \item[The interactions between Ps and laser photons]\mbox{}\\
    Rates of stimulated emissions/absorptions are calculated by Eq. (\ref{eq:B}) for each Ps atom. A laser intensity is approximated as uniform by using the peak value because Ps atoms are assumed to be confined in the small region. 

\end{description}

\section*{References}
\bibliography{ref.bib}

\end{document}